\documentclass{PoS}

\title{Hardware-based Tracking at Trigger Level for ATLAS: The Fast TracKer (FTK) Project}

\ShortTitle{Hardware-based Tracking at Trigger Level for ATLAS: The Fast TracKer (FTK) Project}

\author{\speaker{Johanna Gramling} on behalf of the ATLAS Collaboration\\
       DPNC, Universite de Geneve\\
       E-mail: \email{johanna.gramling@cern.ch}
        }

\abstract{Physics collisions at 13 TeV are expected at the LHC with an average of 40-50 proton-proton collisions per bunch crossing under nominal conditions. Tracking at trigger level is an essential tool to control the rate in high-pileup conditions while maintaining a good efficiency for relevant physics processes. The Fast TracKer is an integral part of the trigger upgrade for the ATLAS detector. For every event passing the Level-1 trigger (at a maximum rate of 100 kHz) the FTK receives data from all the channels of the silicon detectors, providing tracking information to the High Level Trigger in order to ensure a selection robust against pile-up. The FTK performs a hardware-based track reconstruction, using associative memory that is based on the use of a custom chip, designed to perform pattern matching at very high speed. It finds track candidates at low resolution (roads) that seed a full-resolution track fitting done by FPGAs. 
An overview of the FTK system with focus on the pattern matching procedure will be presented. Furthermore, the expected performance and the integration of FTK within the ATLAS trigger system will be discussed.}

\FullConference{The European Physical Society Conference on High Energy Physics\\
		 22-29 July 2015\\
		 Vienna, Austria}

\begin{document}

\section{Motivation}

The Fast TracKer (FTK) Project~\cite{TDR} is an integral part of the trigger upgrade for the ATLAS detector~\cite{ATLAS}. It is designed to provide full tracking information to the ATLAS High Level Trigger (HLT) at the full accept rate of the Level-1 Trigger. 
Including the information provided by FTK in the HLT decisions leads to higher trigger efficiencies for medium-$p_{\rm{T}}$ b's and $\rm{\tau}$'s with high background rejection, since both b-tagging and $\rm{\tau}$ identification rely on track information: b-jets are characterised by a displaced vertex that can be reconstructed from the tracks in the event and jets coming from hadronic $\rm{\tau}$ decays have significantly less tracks in a smaller cone than standard jets~\cite{TDR}. This is especially important for Higgs coupling measurements, where third generation leptons play a crucial role, as well as for SUSY searches, since scenarios with light stops, sbottoms, staus, seem interesting but challenging to rule out or discover~\cite{TDR}.

Furthermore, the primary vertex and the pile-up condition of the event can be determined using FTK tracks. Many trigger algorithms can be improved when including this information in the HLT decision, especially when relying on isolation variables (lepton triggers) and calorimeter information (jet and missing transverse momentum triggers), since they are both affected by pile-up effects.

\section{Main Concepts of FTK}

For every event passing the Level-1 Trigger, FTK performs a hardware-based track reconstruction based on full-precision hit information from all channels of the ATLAS silicon detectors. The resulting tracks are sent to the HLT to be used in the software algorithms. In order to cope with event rates of up to 100 kHz the tracking performed by FTK has to be several orders of magnitude faster than offline tracking. Hence, the processing of the data is organised as parallel as possible: the signals from the detector volume are split into 64 regions, so-called Towers, which are processed independently. Further, the data volume is decreased as much as possible by a custom clustering algorithm defining "hits" which are considered later on instead of the full pixel/strip information. In addition, the hit information is re-binned into coarse-resolution "superstrips" whenever appropriate.

FTK performs the tracking in two steps. 
At first, track candidates are identified by comparing the fired superstrips to predefined trajectories stored in memory. Such a "pattern" refers to a list of superstrips describing the trajectory of a simulated particle as it traverses the detector layers. These track candidates at coarse resolution (roads) seed a full-resolution track fitting done by FPGAs. Only considering hits within these roads reduces the combinatorics significantly and hence makes the fit itself much faster. 
The pattern matching procedure is based on the use of a custom associative memory (AM) chip designed to perform pattern matching at very high speed. It allows to compare the incoming data simultaneously to all stored patterns. 
The parameters of the pattern matching have to be adjusted: while narrow roads permit a fast track fitting, efficient matching requires to store many patterns in the AM. Wide roads, on the other hand, allow for fewer patterns stored but the increased combinatorics within the matched roads slow down the track fitting. This choice is optimised by implementing the feature of variable resolution of the roads via ternary bits in the AM logic~\cite{AM}. Furthermore, the number of matching layers is programmable. 

\section{Data flow within FTK}

\begin{figure}
\centering
\includegraphics[width=0.6\textwidth]{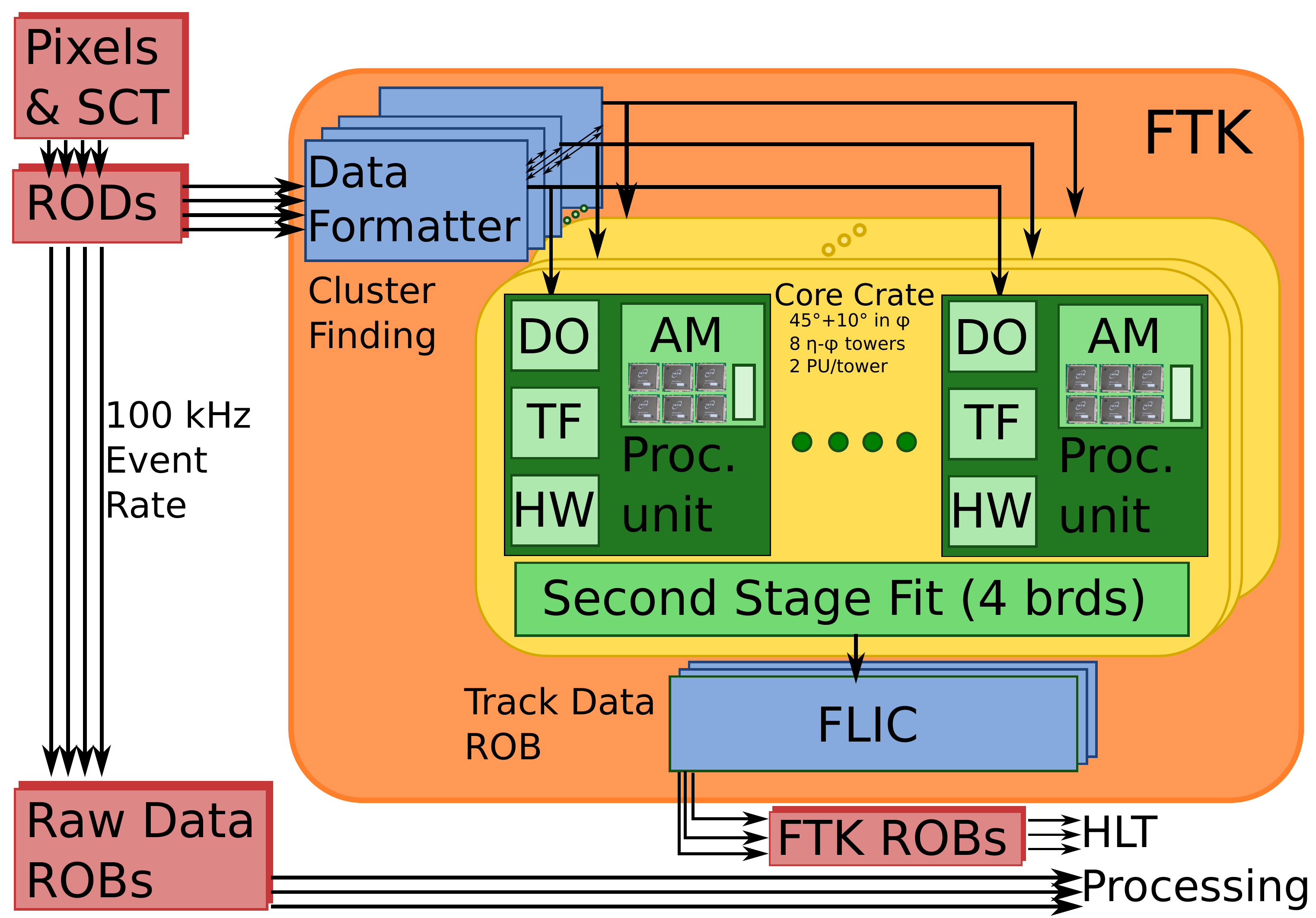}
\caption{Architecture of FTK~\cite{TDR}. AM is the Associative Memory, DO is the Data Organiser, FLIC is the FTK-to-Level-2 Interface Crate, HW is the Hit Warrior, a part of the AUX (Auxiliary Board), ROB is the ATLAS Read Out input Buffer, ROD is a silicon detector Read Out Driver, and TF is the Track Fitter, a part of the AUX. The Second Stage Fit is performed by the Second Stage Board mentioned in the text.}
\label{FTKscheme}
\end{figure}

The data flow and the most important components of FTK are shown schematically in Figure~\ref{FTKscheme}.
Starting from the Input Mezzanine cards (not shown in Figure~\ref{FTKscheme}) which receive data from the tracking detectors and perform the custom pixel/strip clustering algorithm on FPGAs, the hits are sent to the 32 Data Formatters, responsible for the geometrical grouping of the data into the 64 independent towers. 
For each tower, the information is distributed to the corresponding processing unit, which consists of two Auxiliary (AUX) cards and their Associative Memory Board (AMB).
The full-resolution hits are reclustered into coarse superstrips by the AUX Data Organisers that are then communicated to the corresponding AMBs, where the AM chips match the incoming superstrips to the stored track patterns.
The found track candidates are input to the AUX Track Fitter which performs a first tracking within the roads, relying on the full-resolution hit information in eight out of the 12 silicon detector layers.
Subsequently, the AUX Hit Warrior function removes duplicated tracks.
Based on the results, the 32 Second Stage Boards extend the fit to all layers of the Tracker and refine the track parameters in a second fit. Finally, the 2 FLICs take care of the communication with the HLT. In summary, 8 full 9U VME crates and 5 ATCA shelves host about 2000 FPGAs and 8000 custom AM chips, which makes the FTK a - very complex - custom parallel supercomputer.

\section{Expected Performance}
\begin{figure}
\begin{minipage}[t]{0.48\textwidth}
\includegraphics[height=5cm]{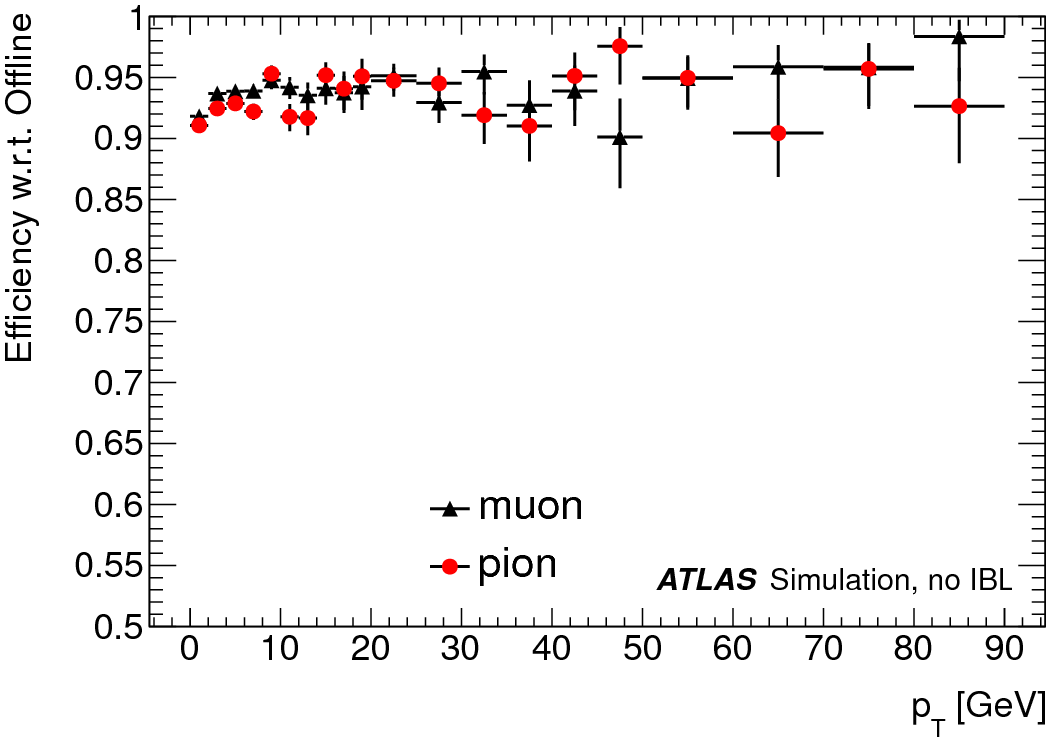}
\caption{Absolute efficiency with respect to truth particles in muon and pion samples versus $p_{\rm{T}}$. The truth particles are required to have a $p_{\rm{T}} > 1\,\rm{GeV}$.\cite{TDR}}
\label{trackeff}
\end{minipage}
\hfill
\begin{minipage}[t]{0.48\textwidth}
\includegraphics[height=5cm]{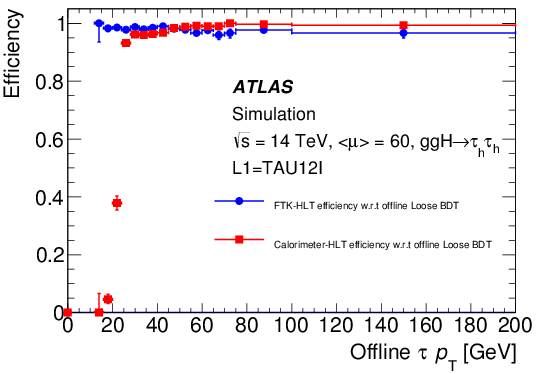}
\caption{ 
$\rm{\tau}$ identification efficiency as a function of offline $\rm{\tau}$ $p_{\rm{T}}$ applying an FTK-based selection (blue) and a calorimeter-based selection (red) at HLT~\cite{pubres}. Efficiency is defined as the fraction of Level-1 $\rm{\tau}$ matched to an offline $\rm{\tau}$\cite{tauID}. 
}
\label{Htt}
\end{minipage}
\end{figure}

Despite the huge increase in speed, the quality of FTK tracks is in many respects comparable to offline tracks. The momentum and angular resolution is only slightly worse and small effects from pile-up enter. Also the b-tagging efficiency is found to be similar to offline in simulation, with high light-jet rejection. The number of vertices found offline and in FTK correspond linearly. As an example, the tracking efficiency versus track $p_{\rm{T}}$ is shown in Figure~\ref{trackeff}. Many more details can be found in the FTK Technical Design Report~\cite{TDR}. With FTK, it is possible to efficiently trigger on 1-prong $\rm{\tau}$'s in a  $H\rightarrow\rm{\tau}\rm{\tau}$ simulation. In particular low-$p_{\rm{T}}$ candidates can be identified and triggered when including FTK tracks in the trigger algorithms. This can be seen in Figure~\ref{Htt}.

\section{Conclusion and Outlook}
The Fast TracKer Project is presented, which aims to provide full tracking information ready for the ATLAS High Level Trigger. This will improve many trigger selections by having precise information about the pile-up condition available for the trigger decision as well as to allow for the design of algorithms to efficiently trigger medium-$p_{\rm{T}}$ b's or $\rm{\tau}$'s. In order to achieve the goal of performing track fitting at a rate of 100 kHz, FTK is built highly parallel and exploits a custom designed AM chip for pattern matching to speed up the track reconstruction.

While Run II of the LHC is ongoing, the first boards of FTK are being installed, already reading ATLAS data in parasitic mode. The full processing of the complete barrel region is expected for spring 2016, after which the complete integration within HLT and the extension to full coverage will follow. In view of the HL-LHC, the FTK concept is discussed to be extended to cope with luminosities higher than the design goal of the current system. Also ideas of using tracking information already at Level 1, and involving an upgraded AM chip are considered.

\end{document}